\documentclass[useAMS,usenatbib]{mn2e}
\usepackage{natbib}
\usepackage{color,graphicx} 
\usepackage{amsmath}        
\usepackage{amsfonts}       
\usepackage{amssymb}        
\usepackage{marvosym}       
\usepackage{wasysym}        
\usepackage{gensymb}        
\usepackage{framed}         
\usepackage{upgreek}

\long\def\symbolfootnote[#1]#2{\begingroup\def\thefootnote{\fnsymbol{footnote}}\footnote[#1]{#2}\endgroup}

\author[Keane et al.]{E.F.~Keane$^{1}$, B.W.~Stappers$^{2}$,
  M.~Kramer$^{1,2}$ \& A.G.~Lyne$^{2}$ \\ $^{1}$ Max Planck Institut
  f\"{u}r Radioastronomie, Auf dem H\"{u}gel 69, 53121 Bonn,
  Germany. \\ $^{2}$ University of Manchester, Jodrell Bank Centre for
  Astrophysics, School of Physics \& Astronomy, Manchester M13 9PL,
  UK.} \date{15 March 2012} \title[A highly-dispersed coherent radio
  burst] {On the origin of a highly-dispersed coherent radio burst}

\begin{document}

\maketitle

\begin{abstract}
  We discuss the possible source of a highly-dispersed radio transient
  discovered in the Parkes Multi-beam Pulsar Survey (PMPS). The pulse
  has a dispersion meausure of $746\;\mathrm{cm}^{-3}\;\mathrm{pc}$, a
  peak flux density of $400$~mJy for the observed pulse width of
  $7.8$~ms, and a flat spectrum across a $288$-MHz band centred on
  $1374$~MHz. The flat spectrum suggests that the pulse did not
  originate from a pulsar, but is consistent with radio-emitting
  magnetar spectra. The non-detection of subsequent bursts constrains
  any possible pulsar period to $\gtrsim1$~s, and the pulse energy
  distribution to being much flatter than typical giant pulse emitting
  pulsars. The burst is also consistent with the radio signal
  theorised from an annihilating mini black hole. Extrapolating the
  PMPS detection rate, provides a limit of
  $\Omega_{BH}\lesssim5\times10^{-14}$ on the density of these
  objects. We investigate the consistency of these two scenarios, plus
  several other possible solutions, as potential explanations to the
  origin of the pulse, as well as for another transient with similar
  properties: the Lorimer Burst.

\end{abstract}


\section{Introduction}

There are many known and proposed sources of transient radio wave
emission in the Universe --- from well-known terrestrial, solar system
and Galactic sources to a menagerie of hypothesised sources covering a
broad range in terms of potential detectability and plausability. Many
of the known sources recur, e.g. solar radio bursts occur every
day~\citep{nglt02} and the pulses from some radio pulsars repeat so
reliably that they can be used as `cosmic clocks' to detect the
stochastic gravitational wave background~\citep{haa+10}. Some bursts
however are less regular in their recurrance, and indeed several radio
transients have never been observed to repeat. This latter group are
of particular interest, especially as a number of the postulated
sources are one-time-only events which would never repeat, and as the
numbers detected should rise dramatically with the onset of the
`all-sky transient monitoring' capabilities of LOFAR~\citep{sha+11},
the VLA~\citep{pcbw11}, MeerKAT~\citep{bbjf09}, ASKAP~\citep{jtb+08}
and ultimately the SKA~\citep{cc11}.

The numerous radio pulsar surveys constitute a rich, and perhaps the
best available, data archive for exploring the parameter space of
potential radio transients --- especially at millisecond to second
scales. 
Many have never been searched for transient bursts.
In one such search of the Parkes Multi-beam Pulsar Survey (PMPS,
\citet{mlc+01}), a highly-dispersed single burst of radio emission was
discovered~\citep{kle+10,kkl+11}. This burst, which was detected, in a
beam with $7$~arcmin half-power radius, in the direction of RA $=$
18:52:05, DEC $=$ $-$08:29:35, i.e. l $=$ $25.4342\degree$, b $=$
$-4.0042\degree$, is the subject of this paper.
In \S~\ref{sec:pulse} we describe the properties of the pulse. In
\S~\ref{sec:solutions} we consider possible solutions as to its
origin.  Throughout, we present the corresponding discussion for the
`Lorimer Burst' (LB from herein) reported by \citet{lbm+07}. In
\S~\ref{sec:conc_disc} we present our conclusions and a discussion on
these matters.

\section{The Pulse}\label{sec:pulse}
The pulse in question was received in the PMPS on June 21st 2001 (MJD
52081) in observation PM0141\_017A1,
i.e. in beam A (the 10th beam, in the outer hexagonal ring of the
receiver) in the 17th pointing recorded on survey tape 141. The
observation started at 12:57:32 UTC (MJD $52081.539953703701$ UTC) and
the pulse was detected $278.795$ seconds into the observation, i.e. at
13:02:10.795 UTC (MJD $52081.543180497682$ UTC). The Simbad
astronomical database lists only 7 sources within a radius of $7$
arcmin: six are optical/IR stars, and one is an X-ray source. There
are no H$\alpha$ or HI anomalies at this position.


\begin{figure}
  \begin{center}
    \includegraphics[trim = 20mm 0mm 0mm 20mm, clip,scale=0.35,angle=0]{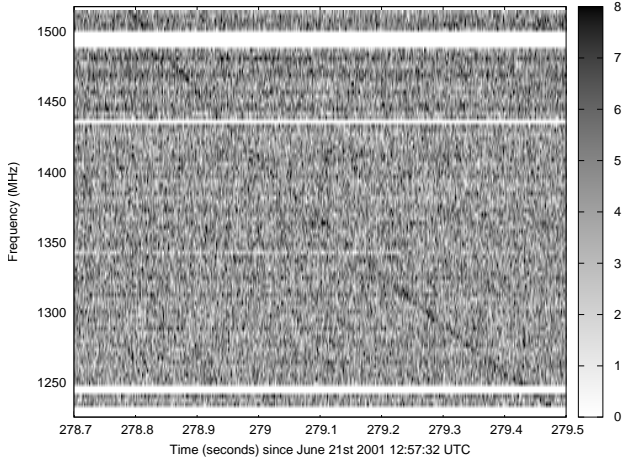}
  \end{center}
  \vspace{-25pt}
  \caption{\small{The pulse from J1852$-$08, as a function of sky
      frequency and observing time, with amplitude represented by a
      grey scale. Here eight $250$-$\upmu$s samples have been added
      together to produce the grey scale (0=white, 8=black) from the
      originally 1-bit data. Nine of the 96 channels contain no
      information and have been set to zero. The frequency-dependent
      delay is proportional to $f^{-2.02\pm0.01}$ and the best-fit
      dispersion measure is $746\;\mathrm{cm}^{-3}\;\mathrm{pc}$.}}
  \label{fig:sweep}
\end{figure}

\textit{Characteristics}: The pulse was detected in only one of the 13
beams of the receiver, with a peak signal-to-noise ratio (S/N) of
$16.3$. Figure \ref{fig:sweep} shows the pulse as a function of
observing time and sky frequency. A frequency-dependent delay is
evident. The dispersion measure (DM) which produces the highest S/N,
when dedispersed with respect to the standard relation for a cold
ionised interstellar medium
$t_{\mathrm{delay,s}}=4150\mathrm{DM}/f_{\mathrm{MHz}}^2$, is
$746\pm1\;\mathrm{cm}^{-3}\;\mathrm{pc}$. Testing how well such a law
is obeyed by fitting for a frequency-dependent delay function which is
proportional to $f^{\alpha}$ gives $\alpha=-2.02\pm0.01$. This is
robust to dividing the 96-channel data
into as many as 16 parts, before obtaining a time-of-arrival (using
standard pulsar timing methods) at each sub-band and then fitting. For
16 sub-bands the average S/N per sub-band is $\approx4$. Finer
frequency resolution results in unacceptably low S/N in the sub-bands.

The observed pulse width, when the data are dedispersed to
$1516.5$~MHz (the top of the band) using a DM of
$746\;\mathrm{cm}^{-3}\;\mathrm{pc}$, is $7.8$~ms, (the PMPS time
resolution is $250\;\upmu$s). The pulse width is $7.7$~ms in the top
half of the band (dedispersed to $1516.5$~MHz) and $8.4$~ms in the
bottom half of the band (dedispersed to $1372.5$~MHz). Finer frequency
resolution shows the pulse width to be constant (within the limits of
what can be determined for the S/N) as a function of frequency across
the entire $288$-MHz band. The pulse is barely resolved (if at all),
as the dispersive smearing time within the $3$-MHz frequency channels
is $7.1$~ms ($9.9$~ms and $5.7$~ms) at the middle (bottom and top) of
the band. Removing this contribution from the observed pulse width
reveals that any intrinsic width, combined with any scatter
broadening, amount to no more than $3$~ms. A steep power law, e.g. the
Kolmogorov scenario with $W\propto f^{-4.4}$ can be ruled out, as can
the predicted scatter broadening time of $\sim 130$~ms from the
empirical estimate of \citet{bcc+04}, although the latter is known to
be uncertain to at least 2 orders of magnitude.

With a measure of S/N as a function of frequency, we can, with a
knowledge of the sensitivity of the instrument, determine the radio
spectrum of the pulse.
This results in peak flux densities, for the \textit{observed} pulse
width, ranging from as low as $360$ to as high as $510$~mJy across the
band, with a slight dip in flux density in the middle of the band (see
Figure~\ref{fig:sweep}).
Within the error in this estimation (expected to be at least $30\%$)
and the limits of the available S/N, this is consistent with a flat
spectrum. Another factor, potentially the dominant source of
uncertainty, is the angular and frequency dependence of the telescope
gain. The pulse is unlikely to have been detected along the central
axis of the beam corresponding to maximum sensitivity, where the gain
is $0.581$~K/Jy~\citep{mlc+01}. The beam responses are well described
as Gaussian, with half-power-beam-widths (HPBWs) of $14.5$~arcmin
at the central frequency of $1374$~MHz. In addition to this angular
dependence the HPBW will scale linearly with frequency so that at the
top of the band, at $1518$~MHz, it will be narrower than at the bottom
of the band, at $1230$~MHz, in proportion to the ratio of these
frequencies. Clearly the spectrum of any incident astrophysical signal
will be made steeper by this effect if detected off-axis.
\citet{bbe+11} discuss empirical measurements of this effect where a
positive intrinsic spectral index $\beta$ can appear as steep as
$\beta-3$. Thus there is an extra (perhaps very large) uncertainty in
the spectrum of the source, and the spectrum is likely to be flatter
than observed.
Below we take the peak flux density of the pulse (at the observed
pulse width) to be $400$~mJy. The intrinsic fluence, $\mathfrak{F}$,
of the $7.8$-ms burst is thus:
\begin{equation}
  \mathfrak{F}= 1.2\times10^{13}
  (D/20\;\mathrm{kpc})^2
  \mathrm{J}\;\mathrm{Hz}^{-1} \;.
\end{equation}

\textit{Distance}: The distance to the source of the pulse, whatever
it may be, is uncertain. The only available estimate is to invert the
relationship between DM and distance:
$DM=\int_0^{D}n_{\mathrm{e}}dl$. This requires a model of the free
electron density. The current best model is NE2001~\citep{cl02} which
predicts that a maximum of $533\;\mathrm{cm}^{-3}\;\mathrm{pc}$ of the
DM is contributed by the Galaxy, towards the total value of
$746\;\mathrm{cm}^{-3}\;\mathrm{pc}$. The six pulsars within a radius
of two degrees of this line of sight have DMs of no more than
$282\;\mathrm{cm}^{-3}\;\mathrm{pc}$. To estimate the distance there
are two options: (i) Assume that the NE2001 model is incorrect along
this line of sight, so that all of the DM contribution is due to
Galactic material, and the source is then within our Galaxy at a
distance of $\lesssim20$~kpc (this being where the `Galaxy ends'
according to the NE2001 model); (ii) Take the NE2001 estimate to be
correct, in which case the excess DM of
$223\;\mathrm{cm}^{-3}\;\mathrm{pc}$ would be due to the intergalactic
medium and any putative host galaxy for the source. In the second
scenario, repeating the analysis of \citet{lbm+07}, an extremely large
distance of $\approx 500h^{-1}$~Mpc results.

\textit{Re-observations}: Deciding between these two possibilities is
crucial as the implied luminosities, and therefore likely progenitors,
depend on it. Motivated by this we performed follow-up observations
using the Parkes Telescope to test the hypothesis that the source is
Galactic (i.e. NE2001 is incorrect) and a pulsar which emits giant
pulses (the most likely Galactic solution). In this scenario many more
weaker pulses would be expected to be easily detected. We used
analogue filterbanks, the same backend as was used in the original
survey~\citep{mlc+01}, with $0.5$-MHz and $100$-$\upmu$s frequency and
time resolution respectively, recording Stokes I with 1-bit
digitisation. Concurrently we recorded Stokes IQUV data, with the
digital filterbanks with 8-bit digitisation, with the aim of looking
more closely at this data, with polarisation information, and with
more dynamic range, if a second event was detected in the total
intensity data. In $15$ hours of follow-up only two events were
detected which were inconsistent with radiometer noise, both of which
were clearly due to narrow-band radio frequency interference (RFI). No
further pulses from J1852$-$08 were seen in a total of $15.5$ hours of
observation. In the following section we investigate the implications
of this.

\section{Possible Solutions}\label{sec:solutions}

\textit{Radio Frequency Interference}: The strongest signals
detectable by any radio telescope are often of terrestrial origin.
In searches for isolated astrophysical bursts there are several steps
we can follow to minimise the effects of RFI. For instance, the
zero-DM subtraction technique can be used to remove broadband RFI,
using the fact that it will be strongest at a DM of
zero~\citep{ekl09}. Narrow-band RFI can be removed by keeping an
account of the bandpass as a function of time and excising anomolously
high or variable channels. Furthermore, for pulsar surveys performed
at Parkes using the 13-beam 21-cm receiver,
boresight sources are not expected to appear in more than 4 beams at
once. Most astrophysical signals are detected in only one beam, and
many RFI signals are detected in several beams, i.e they are detected
in the sidelobes of the beams' gain patterns, so that a multi-beam
coincidence test can be used to discriminate against these undesired
signals. This can take the form of a post-facto comparison of detected
events~\citep{kle+10} or, as has recently been implemented by
\citet{kbb+12}, a full cross-correlation of the signals from each
beam.
The pulse under consideration here survives all of these checks which
leads us to conclude that it is astrophysical in origin. As stated in
\S~\ref{sec:pulse} the pulse follows the theoretical $f^{-2}$
dispersion law without deviation, and is detected in only one beam of
the 21-cm receiver. This clearly contrasts with typical RFI signals,
but also with less commonly encountered RFI signals such as the
``peryton'' signals~\citep{bbe+11,kbb+12}. These signals are detected
in most/all beams, but with atypical frequency-dependent delay. This
delay is somewhat quadratic, but with `kinks' where $dt/df\approx0$.
These authors also noted that the inferred DM for these signals, if
one fits for the theoretical quadratic dispersion law, is, in the case
of 15 of the 21 perytons reported, within $10\%$ of
$375\;\mathrm{cm}^{-3}\;\mathrm{pc}$, the DM for the LB. Other curious
characteristics of the peryton signals include that: they all occured
in the UTC range $0-3$, have a tendency to appear in the latter
$\sim20\%$ of each second, and are separated by gaps of $\sim22$
seconds. In the light of the disimilarity between these signals and
the burst discussed in this paper we are confident that the two
phenomena are not linked. Furthermore we agree with the conclusion of
\citet{bbe+11} that the LB originated from a boresight astrophysical
source, as originally concluded by \citet{lbm+07}.

\textit{Pulsar Giant Pulse}: Pulsar spectra are typically steep: their
mean spectral index is $-1.8$, with values steeper than $-1$ being
very uncommon~\citep{mkk+00}. Nevertheless, despite the flat spectrum
of the pulse, we consider this possibility as, on a pulse-by-pulse
basis pulsar spectra can vary dramatically~\citep{kkg+03}, and because
the radio-emitting magnetars show flat spectra (e.g. \citet{lbb+10}
and references therein). This scenario presupposes that the NE2001
model is sufficiently uncertain, i.e. the free electron content is
underestimated by more than 50\% along the line of sight to the
source. This is
possible; see e.g. \citet{gmcm08} who showed errors of a factor of two
at high Galactic latitutes, and \citet{dtbr09} for a comparison with
VLBI-derived distances. The most likely
Galactic source to consider is that of a `giant pulse' (GP) from a
pulsar at the edge of the Galaxy. GPs are sometimes (arbitrarily)
defined to be those pulses where the pulse `energy' (i.e. the product
of the peak flux density and the effective pulse width) is more than
10 times the mean~\citep{kni06}. There are at least 14 pulsars known
to exhibit GPs, the most notable being the Crab Pulsar, which we will
use as a template below. The pulse energy distributions for GPs follow
power-laws such that the number of GPs with energy
$E>E_{\mathrm{thresh}}$ is proportional to
$E_{\mathrm{thresh}}^{-\alpha}$. The measured value of $\alpha$ in the
most recent analysis of the main pulse GPs of the Crab is
$2.1$~\citep{kss10}. The lowest measured value is $1.4$ for the
millisecond pulsar B1937$+$21~\citep{spb+04}
and the highest is $4.5-4.8$ for the 0.9-s pulsar
B0031$-$07~\citep{ke04}.

The detected pulse had a S/N of $16.3$, whereas we would have been
capable of easily detecting pulses at S/Ns as low as 5. For a
GP-emitting pulsar this means that we might have expected
$(16.3/5)^{\alpha}$ weaker pulses (this is $\sim10$ for the Crab's
$\alpha=2.1$) assuming similar pulse widths, or that the probability
we get a pulse $>5\sigma$ in a given (unknown) pulse period $P$ is
$(16.3/5)^{\alpha}/N$ where $N=(T_{\mathrm{obs}}/P)$ is the total
number of periods observed and $T_{\mathrm{obs}}$ is the observation
time. Thus the probability of not seeing a single weaker pulse for $N$
periods in a row is $(1-(16.3/5)^{\alpha}/N)^N$ which (assuming
$N\gg1$, which it no doubt will be for any reasonable pulsar period)
is just $\exp(-(16.3/5)^{\alpha})$. This has the (perhaps
counter-intuitive) implication that the limit on $\alpha$ is
independent of the duration of the followup, as long as $N\gg1$ and
there are no re-detections. So, a probability of $<10^{-5}$ implies
$\alpha\gtrsim 2$ (such as for the Crab) is ruled out and for a
probability of $<10^{-2}$, the range $\alpha\gtrsim 1.2$ is ruled
out. Very flat distributions with $\alpha\lesssim1$ are not ruled out.

Although the limit on $\alpha$ is no better for the $15.5$ hours of
followup as compared to the 35-minute PMPS observation alone, the
limit on the GP rate is much stronger. The rate per period is
$<6\times10^{-7}(P/P_{\mathrm{Crab}})$, whereas pulses with the same
energy
($400\;\mathrm{mJy}\times7.8\;\mathrm{ms}=3.1\times10^3\;\mathrm{Jy\;\upmu
  s}$) occur with a much higher probability of $\approx2\times10^{-3}$
in the case of the Crab's main-pulse GPs (see the top panel of Figure
5 in \citet{kss10}). However, we should consider the intrinsic
fluence, $SWD^2$, for a fair comparison. Taking $20$~kpc as the
distance to J1852$-$08, and recalling that the distance to the Crab is
$2$~kpc, we see that we must consider the rate of pulses with energy
$3.1\times10^5\;\mathrm{Jy\;\upmu s}$ which is
$2\times10^{-3}(100)^{-2.1}\approx 10^{-7}$ for the Crab. This is not
excluded by our rate limit, however we know $\alpha<1.2$, not $2.1$ as
for the Crab. Also, this implicitly assumes that the GP rate is
identical to the Crab at an intrinsic fluence of
$3.1\times10^5\;\mathrm{Jy\;\upmu s}\;\times (20\;\mathrm{kpc})^2$.
If we consider the rates to be identical at an intrinsic fluence $100$
times lower and utilise our $\alpha$ limit, we get a rate of $\geq
2\times 10^{-3} (100)^{-1.2} \approx 8\times 10^{-6}$. Choosing a
lower intrinsic energy as a reference point gives an even more
constraining limit, although it is unclear where this point should be
chosen.

We can see that a long period pulsar with (say) $P\gtrsim1$~s, or one
with a GP burst rate $10-100$ times less than the Crab,
or, equivalently, a source with a high nulling fraction, is not
ruled out. In the past 6 years a number of long-period pulsars have
been discovered which might fit these criteria. The first group are
the three radio-emitting magnetars. These sources show large
modulation in single pulse flux density,
and flat radio spectra~\citep{ljk+08}. However, unlike the J1852$-$08
pulse, they are seen to be `on' for very long timescales and have been
tracked over several years. The so-called ``RRATs'' emit detectable
pulses at a rate per period of between $10^{-3}-10^{-1}$. It is
unclear, in most cases, whether or not these sources null, but the
lower limit on the ratio of peak to average flux densities can be
$>100$ (e.g \citet{km11}). The pulse energy distributions of these
sources are not yet well studied, although some seem to show
log-normal distributions~\citep{kle+10} whereas the two sources with
power-law distributions show $\alpha=2$ and $3$ respectively (Miller
et al., in prep.). With a single pulse it is impossible to pass
comment on the likelihood of a log-normal (or any other $2$ parameter)
distribution.


We applied the same analysis to the LB, where 40 hours of followup
observation have been reported without a further
detection~\citep{lbm+07}. In this case, we rule out $\alpha\gtrsim
1.0$ with a probability of $10^{-7}$ (and $\alpha\gtrsim 0.5$ with a
probability of $10^{-2}$), where we have adopted Lorimer et al's S/N
of $100$. The limit on the GP rate per period is
$<2\times10^{-7}(P/P_{\mathrm{Crab}})$. In the case of the LB, where
the DM in excess of the NE2001 value is so much larger than in the
J1852$-$08 case, and hence that the distance is apparently $\gg
10$~kpc, this rate can be easily scaled to $1$. However, using the
nominal distance of $500$~Mpc, the inferred radio luminosity is
$\sim8$ orders of magnitude brighter than the brightest pulse ever
detected from the Crab, which seems much too luminous to be due to the
same mechanism. For both bursts, we note that the above calculation of
the limit for $\alpha$ implicitly assumes a rate of 1 per
$T_{\mathrm{obs}}$, whereas the best estimate for the rates, in both
cases, can be taken to be $1.0^{+2.3}_{-0.8}$ per
$T_{\mathrm{obs}}$~\citep{geh86}, with corresponding probability of
$\exp(-X(16.3/5)^{\alpha})$, where $X \in (0.2,3.3)$. If the true rate
were lower than the nominal rate used, the $\alpha$ limit would be
less constraining than stated above, whereas if the rate is in fact
higher the $\alpha$ limit would be even stronger.

\textit{Annihilating Black Holes}: One theorised source of `single
event' radio signals are annihilating mini black holes. Evaporating
black holes with $M_{\mathrm{BH}}<10^{13}$~kg can create
electron-positron pairs (as $kT_{\mathrm{BH}}>2m_{\mathrm{e}}c^2$). If
evaporation can only proceed down to a mass $M_{\mathrm{crit}}$, at
which point the energy of $E=M_{\mathrm{crit}}c^2$ is released, then
pairs with (initial) Lorentz factors of
$\gamma=10^{13}\;\mathrm{kg}/M_{\mathrm{crit}}$ are created. As first
pointed out by \citet{ree77}, such a ``fireball'' of relativistic
pairs, which would have energy $E=10^{30}/\gamma$~J and contain
$E/(\gamma m_{\mathrm{e}}c^2)=10^{43}/\gamma^2$ pairs, expanding into
the surrounding magnetic field of the interstellar medium, will
produce surface currents and a radio burst. \citet{bla77} showed that
the pairs will be sufficiently energetic to do this, while avoiding
annihilation, in the range $10^{5}<\gamma<10^{7}$, and calculated the
energy spectrum of the radio pulse to be:
\begin{equation}
I_{\nu\Omega}=6.7\times10^{-13}\frac{E^{4/3}|F(\nu/\nu_{\mathrm{c}})|^2}{(B\sin\theta)^{2/3}\gamma^{8/3}}\;\mathrm{J}\;\mathrm{Hz}^{-1}\;\mathrm{sr^{-1}}
\; ,
\end{equation}
where $F(\nu/\nu_{\mathrm{c}})$ describes the shape of the spectrum, a
power-law with spectral index $-0.6$ up to $\nu_{\mathrm{c}}$, a
critical frequency $\sim10$~GHz, above which the spectrum steepens
significantly. Substituting in the earlier expression for $E$ and
parameterising $b=(B\sin\theta)/(5\;\mathrm{\upmu G})$, and
$\gamma_{5}=\gamma/10^5$ this becomes:
\begin{equation}
I_{\nu\Omega}=1.1\times10^{14} b_{\mathrm{5\upmu G}}^{-2/3}
\gamma_{\mathrm{5}}^{-4} |F(\nu/\nu_{\mathrm{c}})|^2
\;\mathrm{J}\;\mathrm{Hz}^{-1}\;\mathrm{sr^{-1}} \; .
\end{equation}
Such a pulse occurs in a single radio frequency cycle (and so is
broadband) at a frequency of $\sim1$~GHz. The observed pulse width
$\tau_{\mathrm{obs}}$ is smeared with respect to the intrinsic width,
so that the \textit{inferred} `radio pseudo-luminosity' of the radio
pulse $L\approx(4\pi I_{\nu\Omega})/\tau_{\mathrm{obs}}$ is simply:
\begin{equation}
L_{\mathrm{BH}}\approx\frac{140}{\tau_{\mathrm{obs}}}
b_{\mathrm{5\upmu G}}^{-2/3} \gamma_{\mathrm{5}}^{-4}
|F(\nu/\nu_{\mathrm{c}})|^2 \;\mathrm{Jy}\;\mathrm{kpc}^{2} \; .
\end{equation}
At $\nu/\nu_{\mathrm{c}}\approx0.1$,
$|F(\nu/\nu_{\mathrm{c}})|^2\approx 1$, so in our parameterisation we
have $L_{\mathrm{BH}}\sim140/\tau_{\mathrm{obs}}$. But we know that,
by definition, $L=SD^2$, where $S$ is flux density and $D$ is
distance. Thus, assuming the burst under consideration here is due to
such an annihilating black hole we can infer a distance from the
observed $S$, $\tau_{\mathrm{obs}}$ and expected $L_{\mathrm{BH}}$
value. This yields a distance of $\lesssim200$~kpc. Larger values of
$\gamma_5$ or the lower sensitivity of an off-axis detection both
lower the distance. In theory, this estimate for distance in the black
hole scenario can be compared with the DM estimate for the distance as
a check of consistency. We again have two scenarios: (i) if the NE2001
model is correct then the distance is much too short and inconsistent
with the DM distance; (ii) if the NE2001 model is sufficiently
incorrect, so that we can attribute essentially all the dispersion to
the $\sim20$~kpc of Galactic material along the line of sight, the
scenario is not inconsistent. In the case of the LB, repeating this
analysis we infer a distance of $\lesssim 25$~kpc, which is,
regardless of whether or not NE2001 is hugely incorrect along that
particular line of sight, much too short. Thus this scenario is not a
consistent explanation for the LB.


\textit{Other Solutions}: In double neutron star systems, one star
will be recycled to millisecond periods, whereas the second will have
a longer period (e.g. \citet{lor08} and references
therein). \citet{hl01b} consider such a binary, on the verge of
merging due to the emission of gravitational waves, where the longer
period star has a magnetic field of $\sim10^{15}$~G and a period of
$\gtrsim 10$~s. The millisecond pulsar fuels a flow of plasma in the
light cylinder of the long-period pulsar, within which it is
completely enclosed, resulting in the loss of orbital and spin energy
to a broadband coherent millisecond radio burst. The flux density
detected at Earth for such an event, occuring at a distance $D$, is
$S\approx 1\;\mathrm{mJy}(\epsilon/0.1)(100\;\mathrm{Mpc}/D)^2
B_{15}^{2/3}$ where $\epsilon$ is an efficiency factor, and $B_{15}$
is the magnetic field strength of the long period neutron star in
units of $10^{15}$~G. Applying this to the pulse under consideration
here implies a distance of $\sim5$~Mpc ($\sim 500$~kpc) for magnetic
field strengths of $10^{15}$~G ($10^{12}$~G), which is not
inconsistent with the observed dispersion, for a sufficiently
incorrect NE2001 estimate along this line of sight. In the case of the
LB, this yields a distance of $\sim600$~kpc ($\sim60$~kpc), which is
too small a distance to be consistent, regardless of the corrrectness
of NE2001 for the line of sight. However, it is highly uncertain as to
whether the signal could propagate through the ``plasma shroud'' of
the system, and the volumetric merger rate suggests that these events
are unlikely to occur at distances $\ll 100$~Mpc~\citep{hl01b}. A
confirmation would neccessarily require a detection of the
gravitational wave counterpart signal, but as both pulses under
discussion here occured before the LIGO and GEO600 detectors came
online, no such check is possible.

\citet{cn71} and \citet{col75} considered the case of a supernova
shell expanding into the magnetic field of the pre-existing star. The
shell ``combs'' the magnetic field into the radial direction,
producing a current sheet and an associated coherent radio burst. As
for the merger scenario, given the expected rates, the distance such
events are expected to occur at is $\sim 100$~Mpc. Furthermore, for a
volumetric supernova rate of
$9\times10^{-5}\;\mathrm{Mpc}^{-3}\;\mathrm{y}^{-1}$~\citep{hbk+11},
the number of supernovae expected in the PMPS would be just $\sim0.1$
assuming it would be sensitive out to $500$~Mpc in all directions,
which itself is highly optimistic given the effects of dispersion and
scattering on sensitivity. The predicted allowable pulse energies
cover a large range: $10^{27}-10^{35}$~J, with the 1970s limits just
probing the upper end of this range for sources at
100~Mpc~\citep{mc78}. Taking the pulse under consideration here to be
due to such an event the implied energy at $100$~Mpc is
$\approx10^{29}$~J. The same calculation for the LB gives $\sim 5
\times 10^{30}$~J for the same distance. The wide (and therefore
unconstraining) range of allowable energies mean that a distance
consistent with the dispersion can easily be found.

Given the wide range of Lorentz factors possible ($\sim10-10^6$), one
might also devise a relativistic source to fit the observed distance
and flux density, e.g. a precessing jet from a microquasar
(R. Spencer, private communication). All of the other known types of
radio transient signals can be dismissed due to either their
timescales, their lack of dispersion (as nearby sources), or
both~\citep{kea10a}.

\section{Conclusions \& Discussion}\label{sec:conc_disc}
The pulse discussed here is consistent with a number of scenarios. It
is consistent with a `giant pulse' from either a young pulsar with a
burst rate perhaps $10-100$ times less than the Crab, or a long period
pulsar, albeit with quite a steep cumulative pulse energy distribution
with $\alpha\lesssim 1$, and a flat radio spectrum more like those of
the radio-emitting magnetars. A second scenario involving the radio
signal from an annihilating mini black hole is also consistent. It is
interesting that the two possibilities can only apply if the NE2001
model is sufficiently incorrect along the line of sight to the source
that all of the dispersion is due to the Galaxy. It is also noteworthy
that neither of these two scenarios give a consistent solution for the
LB, regardless of NE2001's precision. We can use the lack of other
detections in the PMPS to set an upper limit on
$\Omega_{\mathrm{BH}}$, the cosmological density of mini black holes
which produce radio bursts (i.e. primordial black holes within a
certain mass range). From one event within a (say) $20$~kpc radius,
detected during the PMPS, whose duration was $0.21$~y and
field-of-view was $0.55\;\mathrm{deg}^2$, we can extrapolate to obtain
a limit of $\Omega_{\mathrm{BH}}\lesssim 5\times
10^{-14}/\gamma_5$. Although, amongst other things, this assumes that
all such black holes produce radio bursts at the end of their lives,
the implication is that annihilating black holes are an insignificant
contribution to the matter density of the Universe.

Next we considered radio bursts from NS-NS mergers, which, although
the rate is uncertain, can be consistent with an extragalactic source
for the pulse, although again not for the LB. A burst associated with
an expanding supernova shell allows solutions for both pulses,
although the predicted energy range for the pulse (which spans 8
orders of magnitude) means a meaningful comparison with DM-derved
distances is not possible. An important point is that the intrinsic
timescale is unknown for the pulses discussed here. In both cases the
pulse widths are of the order of, and just slightly larger than, the
dispersion smearing time within a single $3$-MHz frequency channel. We
cannot account for any contribution due to scattering in the
interverning medium, as our knowledge of this effect along specific
lines of sight through the Galaxy is very poor. However, this could be
used to decide between the consistent solutions. For instance, if
there is zero (or very little) scattering then we would know that the
intrinsic pulse timescale is $\sim1$~ms and the annihilating mini
black hole (and supernova) scenarios would be ruled out. All we can
say is that for the pulsars closest to the line of sight monitored by
the Lovell Telescope it is difficult to decipher scattering from
intrinsic profile features, although the recent successful work by
\citet{has+12} in this area provides cause for optimism. Transient
signals detected in the future with LOFAR, and other next generation
wide field-of-view telescopes, where rapid localisation,
classification, multi-wavelength followup, and the detection of any
associated gravitational wave signal (e.g. with Advanced LIGO) will be
possible, lead us to believe that unambiguous identification of the
sources of such energetic events will become routine in the SKA era.

\section*{Acknowledgments}
EK acknowledges the FSM for support. The authors thank R. P. Eatough
and the anonymous referee for useful discussion and helpful comments
which have improved the quality of this paper.


\begin{thebibliography}{}

\bibitem[\protect\citeauthoryear{{Bhat} et~al.}{{Bhat} et~al.}{2004}]{bcc+04}
{Bhat} N.~D.~R., {Cordes} J.~M., {Camilo} F., {Nice} D.~J.,  {Lorimer} D.~R.,
  2004, ApJ, 605, 759

\bibitem[\protect\citeauthoryear{{Blandford}}{{Blandford}}{1977}]{bla77}
{Blandford} R.~D., 1977, MNRAS, 181, 489

\bibitem[\protect\citeauthoryear{{Booth} et~al.}{{Booth} et~al.}{2009}]{bbjf09}
{Booth} R.~S., {de Blok} W.~J.~G., {Jonas} J.~L.,  {Fanaroff} B., 2009, (astro-ph/0910.2935)

\bibitem[\protect\citeauthoryear{{Burke-Spolaor} \& {Bailes}}{{Burke-Spolaor}
  \& {Bailes}}{2010}]{bb10}
{Burke-Spolaor} S.,  {Bailes} M., 2010, MNRAS, 402, 855

\bibitem[\protect\citeauthoryear{{Burke-Spolaor} et~al.}{{Burke-Spolaor}
  et~al.}{2011}]{bbe+11}
{Burke-Spolaor} S., {Bailes} M., {Ekers} R., {Macquart} J.,  {Crawford} F.,
  III, 2011, ApJ, 727, 18


\bibitem[\protect\citeauthoryear{{Colegate} \& {Clarke}}{{Colegate} \&
  {Clarke}}{2011}]{cc11}
{Colegate} T.~M.,  {Clarke} N., 2011, PASA, 28, 299

\bibitem[\protect\citeauthoryear{{Colgate} \& {Noerdlinger}}{{Colgate} \&
  {Noerdlinger}}{1971}]{cn71}
{Colgate} S.~A.,  {Noerdlinger} P.~D., 1971, ApJ, 165, 509

\bibitem[\protect\citeauthoryear{{Colgate}}{{Colgate}}{1975}]{col75}
{Colgate} S.~A., 1975, ApJ, 198, 439

\bibitem[\protect\citeauthoryear{{Cordes} \& {Lazio}}{{Cordes} \&
  {Lazio}}{2002}]{cl02}
{Cordes} J.~M.,  {Lazio} T.~J.~W., 2002 (astro-ph/0207156)


\bibitem[\protect\citeauthoryear{{Deller} et~al.}{{Deller}
  et~al.}{2009}]{dtbr09}
{Deller} A.~T., {Tingay} S.~J., {Bailes} M.,  {Reynolds} J.~E., 2009, ApJ, 701,
  1243

\bibitem[\protect\citeauthoryear{{Eatough}, {Keane}, \& {Lyne}}{{Eatough}
  et~al.}{2009}]{ekl09}
{Eatough} R.~P., {Keane} E.~F.,  {Lyne} A.~G., 2009, MNRAS, 395, 410


\bibitem[\protect\citeauthoryear{{Gaensler} et~al.}{{Gaensler}
  et~al.}{2008}]{gmcm08}
{Gaensler} B.~M., {Madsen} G.~J., {Chatterjee} S.,  {Mao} S.~A., 2008, PASA,
  25, 184

\bibitem[\protect\citeauthoryear{{Gehrels}}{{Gehrels}}{1986}]{geh86}
{Gehrels} N., 1986, ApJ, 303, 336

\bibitem[\protect\citeauthoryear{{Hansen} \& {Lyutikov}}{{Hansen} \&
  {Lyutikov}}{2001}]{hl01b}
{Hansen} B.~M.~S.,  {Lyutikov} M., 2001, MNRAS, 322, 695

\bibitem[\protect\citeauthoryear{{Hassall} et~al.}{{Hassall} et~al.}{2012}]
{has+12}
{Hassall} T. et~al., 2012, A\&A, in press (astro-ph/1204.3864).


\bibitem[\protect\citeauthoryear{{Hobbs} et~al.}{{Hobbs}
  et~al.}{2010}]{haa+10}
{Hobbs} G. et~al., 2010, CQG, 27, 084013

\bibitem[\protect\citeauthoryear{{Horiuchi} et~al.}{{Horiuchi} et~al.}
 {2011}]{hbk+11}
{Horiuchi} S. et~al., 2011, ApJ, 738, 154

\bibitem[\protect\citeauthoryear{{Johnston} et~al.}{{Johnston}
  et~al.}{2008}]{jtb+08}
{Johnston} S. et~al., 2008, Experimental Astronomy, 22, 151

\bibitem[\protect\citeauthoryear{{Karuppusamy}, {Stappers}, \& {van
  Straten}}{{Karuppusamy} et~al.}{2010}]{kss10}
{Karuppusamy} R., {Stappers} B.~W.,  {van Straten} W., 2010, A\&A, 515, A36

\bibitem[\protect\citeauthoryear{{Keane}}{{Keane}}{2010}]{kea10a}
{Keane} E.~F., 2010, Ph.D. thesis, University of Manchester

\bibitem[\protect\citeauthoryear{{Keane} et~al.}{{Keane} et~al.}{2010}]{kle+10}
{Keane} E.~F., et al., 2010, MNRAS, 1057

\bibitem[\protect\citeauthoryear{{Keane} et~al.}{{Keane} et~al.}{2011}]{kkl+11}
{Keane} E.~F., {Kramer} M., {Lyne} A.~G., {Stappers} B.~W., {McLaughlin}
  M.~A., 2011, MNRAS, 838


\bibitem[\protect\citeauthoryear{{Keane} \& {McLaughlin}}{{Keane} \&
  {McLaughlin}}{2011}]{km11}
{Keane} E.~F.,  {McLaughlin} M.~A., 2011, Bulletin of the Astronomical Society
  of India, 39, 333

\bibitem[\protect\citeauthoryear{{Knight}}{{Knight}}{2006}]{kni06}
{Knight} H.~S., 2006, Chinese Journal of Astronomy and Astrophysics Supplement,
  6, 41

\bibitem[\protect\citeauthoryear{{Kocz} et~al.}{{Kocz} et~al.}{2012}]{kbb+12}
{Kocz} J., {Bailes} M., {Barnes} D., {Burke-Spolaor} S.,  {Levin} L., 2012,
  MNRAS, 271

\bibitem[\protect\citeauthoryear{{Kramer} et~al.}{{Kramer} 
  et~al.}{2003}]{kkg+03}
{Kramer}, M., et~al., 2003, A\&A, 407, 655

\bibitem[\protect\citeauthoryear{{Kuzmin} \& {Ershov}}{{Kuzmin} \&
  {Ershov}}{2004}]{ke04}
{Kuzmin} A.~D.,  {Ershov} A.~A., 2004, A\&A, 427, 575

\bibitem[\protect\citeauthoryear{{Lazaridis} et~al.}{{Lazaridis} 
  et~al.}{2008}]{ljk+08}
{Lazaridis}, K., et~al., 2008, MNRAS, 390, 839

\bibitem[\protect\citeauthoryear{{Levin} et~al.}{{Levin} 
  et~al.}{2010}]{lbb+10}
{Levin}, L., et~al., 2010, ApJL, 721, L33

\bibitem[\protect\citeauthoryear{{Lorimer}}{{Lorimer}}{2008}]{lor08}
{Lorimer} D.~R., 2008, Living Reviews in Relativity, 11, 8

\bibitem[\protect\citeauthoryear{{Lorimer} et~al.}{{Lorimer}
  et~al.}{2007}]{lbm+07}
{Lorimer} D.~R., {Bailes} M., {McLaughlin} M.~A., {Narkevic} D.~J.,  {Crawford}
  F., 2007, Science, 318, 777


\bibitem[\protect\citeauthoryear{Manchester et~al.}{Manchester
  et~al.}{2001}]{mlc+01}
Manchester R.~N. et~al., 2001, MNRAS, 328, 17

\bibitem[\protect\citeauthoryear{Maron et.~al.}{Maron
  et~al.}{2000}]{mkk+00}
{Maron} O. et~al., {Kijak}, J., {Kramer}, M., {Wielebinski}, R., 2000, A\&A, 147, 195


\bibitem[\protect\citeauthoryear{{Meikle} \& {Colgate}}{{Meikle} \&
  {Colgate}}{1978}]{mc78}
{Meikle} W.~P.~S.,  {Colgate} S.~A., 1978, ApJ, 220, 1076

\bibitem[\protect\citeauthoryear{{Nita} et~al.}{{Nita} et~al.}{2002}]{nglt02}
{Nita} G.~M., {Gary} D.~E., {Lanzerotti} L.~J.,  {Thomson} D.~J., 2002, ApJ,
  570, 423

\bibitem[\protect\citeauthoryear{{Perley} et~al.}{{Perley}
  et~al.}{2011}]{pcbw11}
{Perley} R.~A., {Chandler} C.~J., {Butler} B.~J.,  {Wrobel} J.~M., 2011, ApJL,
  739, L1

\bibitem[\protect\citeauthoryear{{Rees}}{{Rees}}{1977}]{ree77}
{Rees} M.~J., 1977, Nature, 266, 333

\bibitem[\protect\citeauthoryear{{Soglasnov} et~al.}{{Soglasnov}
  et~al.}{2004}]{spb+04}
{Soglasnov} V.~A., et al., 2004, ApJ, 616, 439 

\bibitem[\protect\citeauthoryear{{Stappers} et~al.}{{Stappers}
  et~al.}{2011}]{sha+11}
{Stappers} B.~W. et~al., 2011, A\&A, 530, A80

\bibitem[\protect\citeauthoryear{Staveley-Smith et~al.}{Staveley-Smith
  et~al.}{1996}]{swb+96}
Staveley-Smith L. et~al., 1996, Proc. Astr. Soc. Aust., 13, 243

\end{thebibliography}
\bibliographystyle{mnras}

\end{document}